\documentclass[prb,twocolumn,superscriptaddress, showpacs,floatfix]{revtex4}

\usepackage[dvips]{graphicx}
\usepackage{amssymb, amsmath, bbm}

\def\be{\begin{eqnarray}}
\def\ee{\end{eqnarray}}

\pacs{75.10.Jm, 75.10.Pq}

\begin{document}
\title{Phase diagram  of  a frustrated   mixed-spin ladder with 
diagonal exchange bonds}
\author{N. B.  Ivanov}
\affiliation{ Theoretische Physik II, Universit\"at Augsburg,
D-86135 Augsburg, Germany}
\altaffiliation[Permanent address:]{ Institute of Solid State Physics, 
Bulgarian Academy of
Sciences, Tsarigradsko chaussee 72, 1784 Sofia, Bulgaria}
\author{J. Richter}
\affiliation{Institut f\"ur Theoretische Physik, Universit\"at Magdeburg,
PF 4120, D-39016 Magdeburg, Germany}
\date{\today}
\begin{abstract}
Using  exact numerical diagonalization and the conformal field theory approach,, 
we study the effect of magnetic frustrations due to diagonal exchange 
bonds in a system of two coupled  mixed-spin  
$(1,\frac{1}{2})$  Heisenberg chains. It is established that
relatively moderate frustrations are able to destroy the 
ferrimagnetic state and to stabilize the  critical spin-liquid
phase typical for half-integer-spin antiferromagnetic Heisenberg chains.  
Both phases are separated by a narrow  but finite region occupied by a 
critical partially-polarized ferromagnetic phase. 
\end{abstract}
\maketitle
\section{Introduction}
Many experiments on bimetallic quasi-one-dimensional
(1D) molecular magnets  imply that the  magnetic properties 
of these compounds are basically described by the   Heisenberg spin model 
with antiferromagnetic exchange couplings.~\cite{kahn,hagiwara} 
In  the last  decade, there has  been  an increasing 
experimental and theoretical interest in these mixed-spin systems 
exhibiting  intriguing quantum spin phases and  thermodynamic properties. 
In particular, a number of  recent studies has  been focused on ground-state 
properties of mixed-spin ladders.~\cite{senechal}  

\begin{figure}[hbt]
\samepage
\begin{center}
\includegraphics[width=2.2in]{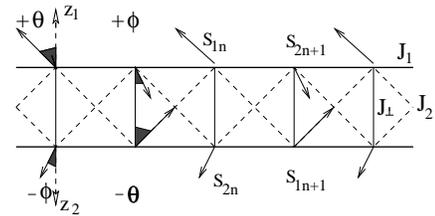}
\vspace{-.4cm}
\caption{\label{ladder} Sketch of the classical canted
state for $J_{\perp}=J_1$ described by the angles 
$(\pm\theta,\pm\phi)$ for every   magnetic cell 
composed of two neighboring rungs.  $\theta$ and $\phi$ 
measure the  deviations of the classical 
spins  from  the ferrimagnetic configuration: 
${{\bf s}_{1}}_n\| {\bf z}_1$ and ${{\bf s}_{2}}_n\| {\bf z}_2$. 
}
\vspace{-.7cm}
\end{center}
\end{figure}

In this paper we study
the ground-state phase diagram of a  ferrimagnetic two-leg
ladder containing frustrating antiferromagnetic diagonal exchange bonds
(see Fig.~\ref{ladder}). The model is defined by the Hamiltonian       
\begin{multline}\label{h}
{\cal H}=\sum_{n=1}^L\left[ 
J_1\left({{\bf s}_{1}}_n\cdot {{\bf s}_{2}}_{n+1}
+{{\bf s}_{2}}_n\cdot {{\bf s}_{1}}_{n+1}\right)
+J_{\perp}{{\bf s}_{1}}_n\cdot {{\bf s}_{2}}_{n}\right]\\
+J_2\sum_{n=1}^L 
\left({{\bf s}_{1}}_{n}\cdot {{\bf s}_{1}}_{n+1}+ 
{{\bf s}_{2}}_{n}\cdot {{\bf s}_{2}}_{n+1}\right) \, ,
\end{multline}
where $L$ is the number of rungs and  the spin operators
${{\bf s}_{1}}_n$ and  
${{\bf s}_{2}}_n$ are defined on the rung with index $n$:  
$\left({{\bf s}_{1}}_n\right)^2=s_1(s_1+1)$, 
$\left({{\bf s}_{2}}_n\right)^2=s_2(s_2+1)$,
$s_1>s_2$, and $J_1,J_2,J_{\perp}>0$. 
We introduce the  frustration parameter   
$\alpha = J_2/J_1$,   the spin ratio $\sigma = s_1/s_2$,
and set the energy and length scales by $J_1\equiv 1$ and $a_0=1$,
where $a_0$ is the spacing between neighboring rungs. In the remainder of this
paper, if not especially noted, we will consider the case $J_{\perp}=J_1$.

As a function of the frustration parameter $\alpha$,
the classical phase diagram of (\ref{h}) exhibits three phases
described by the angles $(\theta,\phi)$ fixing the directions
of the classical spins ${{\bf s}_{1}}_n$ and ${{\bf s}_{2}}_n$ 
in respect to the classical ferrimagnetic configuration 
with spins ${{\bf s}_{1}}_n$ and ${{\bf s}_{2}}_n$ oriented along
the axes ${\bf z}_1$ and ${\bf z}_2$, respectively (Fig.~\ref{ladder}).
The classical canted state (C) shown in Fig.~\ref{ladder} is stable in the
interval $\alpha_{c1}<\alpha<\alpha_{c2}$ where 
$\alpha_{c1}=3[(\sigma^2+1)-\sqrt{(\sigma^2+1)^2-32\sigma^2/9}]/8\sigma$ and
$\alpha_{c2}=[-(\sigma^2+1)+\sqrt{(\sigma^2+1)^2+32\sigma^2}]/8\sigma$ 
are second-order phase transition points separating the C phase from
the ferrimagnetic (F) $(\theta,\phi)=(0,0)$ and the mixed-spin collinear  
$(\theta,\phi)=(\pi/2,\pi/2)$ phases, respectively. In the case of special
interest $(s_1,s_2)=(1,\frac{1}{2})$, the classical transition points are
$\alpha_{c1}=0.3219$
and $\alpha_{c2}=0.4606$. Notice that the square-lattice $J_1-J_2$ mixed-spin Heisenberg
model\cite{ivanov1} exhibits similar classical magnetic phases which persist in the
quantum phase diagram. On the other hand,
the  following analysis of (\ref{h}) implies that in one space
dimension  the ferrimagnetic
phase continues to exist  in the quantum phase diagram, 
whereas the classical collinear magnetic 
state is completely  destroyed by quantum fluctuations.  Instead, for
$\alpha>\alpha_{2c}$ there appears a singlet quantum paramagnetic 
phase which is critical (gapped) for half-integer (integer)
rung spins ${\bf S}_n={{\bf s}_{1}}_{n}+{{\bf s}_{2}}_{n}$.   
As to the classical canted phase, it is argued that 
the longitudinal ferromagnetic order survives quantum fluctuations.
On the other hand, on general grounds we may expect that 
the transverse magnetic ordering does not survive quantum fluctuations.
A closely related phase diagram appears in a special class of lattice
models  with quantum-rotor degrees of freedom.\cite{sachdev}

The following analysis of the quantum phase diagram  is performed
by using exact numerical diagonalization (ED) of small periodic systems,
finite-size analysis of the ED data, and analytical spin-wave calculations. 
The emphasis is on the properties of the quantum paramagnetic phase.
\section{Magnetic phases}
\begin{figure}[hbt]
\samepage
\begin{center}
\includegraphics[width=7cm]{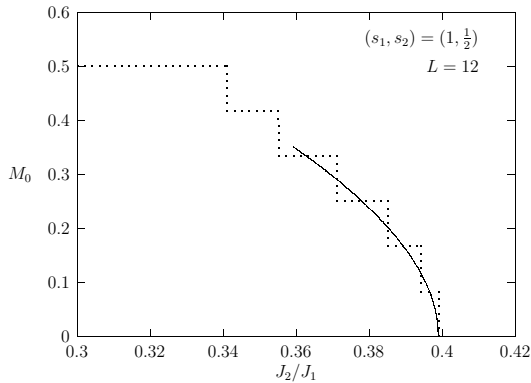}
\caption{\label{fig:m0} ED results for the ferromagnetic moment per rung 
$M_0$ $vs$  $J_2/J_1$ (dashed line). The step-like form of $M_0(J_2/J_1)$
is connected to finite-size effects. The midpoints of
the steps close to $\alpha_{c2}$ are well approximated by the ansatz
$M_0=m_1(\alpha_{c2}-\alpha)^{\beta}+m_2(\alpha_{c2}-\alpha)$, where
$m_1=1.65,m_2=0.56$, and $\beta=\frac{1}{2}$ (solid line).
}
\vspace{-.7cm}
\end{center}
\end{figure}

As may be expected, the positions of the classical phase transition 
points $\alpha_{c1}$ and  
$\alpha_{c2}$ are changed by quantum spin fluctuations. Using ED and a simple
finite-size scaling, it is possible to find precise estimates for the quantum
transition points.
The latter are connected with  the following changes in the total 
spin of the ground state $S_T$: $S_T=(s_1-s_2)L$ for $0\leq \alpha
<\alpha_{c1}$ (F phase),
$0<S_T<(s_1-s_2)L$ for $\alpha_{c1}< \alpha<\alpha_{c2}$ (C phase),
and $S_T=0$ for $\alpha>\alpha_{c2}$ (quantum paramagnetic  phase). 
The extrapolated
data for $L=8,10,12$, and $14$ give the results $\alpha_{c1}=0.341$ and 
$\alpha_{c2}=0.399$ showing that the region occupied by the quantum C phase is
narrowed but definitely finite. Figure~\ref{fig:m0} provides a summary 
of the reported results in
terms of the net ferromagnetic moment per rung $M_0$ for $L=12$. 

The  ferrimagnetic phase has 
already been studied for the model (\ref{h}) without
frustration.\cite{ivanov2}
As $M_0>0$, both magnetic phases (F and C) 
are characterized by quadratic  spin-wave excitations
\be
E(k)=\frac{\rho_s}{M_0}k^2+{\cal O}(k^4)\, ,  
\ee
where $\rho_s$ is the ferromagnetic spin-stiffness constant.\cite{halperin}    
In  approaching $\alpha_{c1}$ from the ferrimagnetic phase, the linear
spin-wave theory predicts that the lower spin wave
branch softens in the vicinity of  $k=\pi$ and the gap at this point
vanishes for $\alpha >\alpha_{c1}$. 
Thus the linear  Goldstone mode,  characteristic of
the classical C phase, seems to survive quantum fluctuations  
although on general grounds it may be expected that the  spin rotation 
symmetry $U(1)$ in the  $xy$ plane is restored in one space dimension, i.e.,  
$\langle {s_1}_n^x\rangle =\langle {s_2}_n^x\rangle= 0$.~\cite{note1}
This scenario is  supported by the renormalization-group analysis
of similar phases in quantum rotor models,\cite{sachdev} implying that
the true tranverse long-range magnetic order in the classical C 
phase is transformed to a quasi-long-range $xy$ order in the quantum
system. On the other hand, the spin-stiffness constant $\rho_s$ remains  
finite in both magnetic phases as well as at the transition point $\alpha_{c1}$. 
Following the terminology of Ref.~\onlinecite{sachdev}, the quantum C phase 
may be called {\em partially polarized ferromagnetic phase}, as
the ferromagnetic moment $M_0$ is less than the maximal value $s_1-s_2$
in the ferrimagnetic phase
(see Fig.~\ref{fig:m0}). This quantum state may also be classified as a
kind of ferromagnetic Luttinger liquid.\cite{bartosch} We postpone
the detail analysis  of this quantum phase for future studies.    
\section{Quantum paramagnetic phase}
Now, let us turn to the region $\alpha >\alpha_{c2}$  of the phase diagram
characterized by $S_T=0$. It is instructive to rewrite (\ref{h}) in the 
following form 
\begin{multline}\label{h1}
{\cal H}=\sum_{n=1}^L\left(J_1{\bf S}_n\cdot {\bf S}_{n+1}
+J_{\perp}{{\bf s}_{1}}_n\cdot {{\bf s}_{2}}_n\right)\\
+(J_2-J_1)\sum_{n=1}^L 
\left({{\bf s}_{1}}_{n}\cdot {{\bf s}_{1}}_{n+1}+ 
{{\bf s}_{2}}_{n}\cdot {{\bf s}_{2}}_{n+1}\right) \, .
\end{multline}
The operator $2{{\bf s}_{1}}_n\cdot {{\bf s}_{2}}_n=
\left({\bf S}_n\right)^2-s_1(s_1+1)-s_2(s_2+1)$ is a conserved quantity
for  $J_2=J_1$ and  at this special point
 the low-energy physics  of (\ref{h}) is described by the antiferromagnetic 
spin-$S$ Heisenberg chain:
$\left({\bf S}_n\right)^2=S(S+1), S=s_1-s_2,s_1-s_2+1,\ldots,s_1+s_2$.
In the case of special interest $(s_1,s_2)=(1,\frac{1}{2})$, and for 
relatively small interchain couplings $J_{\perp}\lessapprox
1.59J_1$, we have numerically found   that  all the rung spins
are characterized by  $S=\frac{3}{2}$, so that in the low-energy sector
the ladder model (\ref{h}) is equivalent 
to the $S=\frac{3}{2}$ antiferromagnetic Heisenberg chain. 
\begin{figure}[hbt]
\samepage
\begin{center}
\includegraphics[width=7cm]{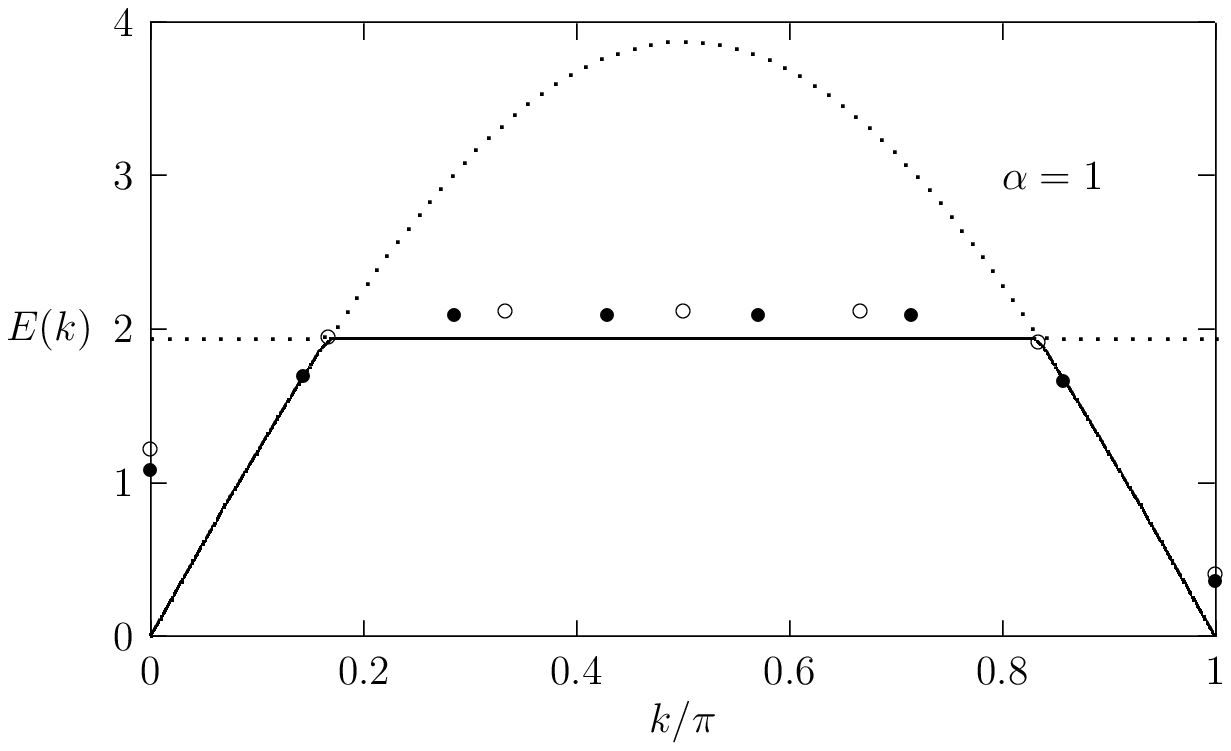}
\includegraphics[width=7cm]{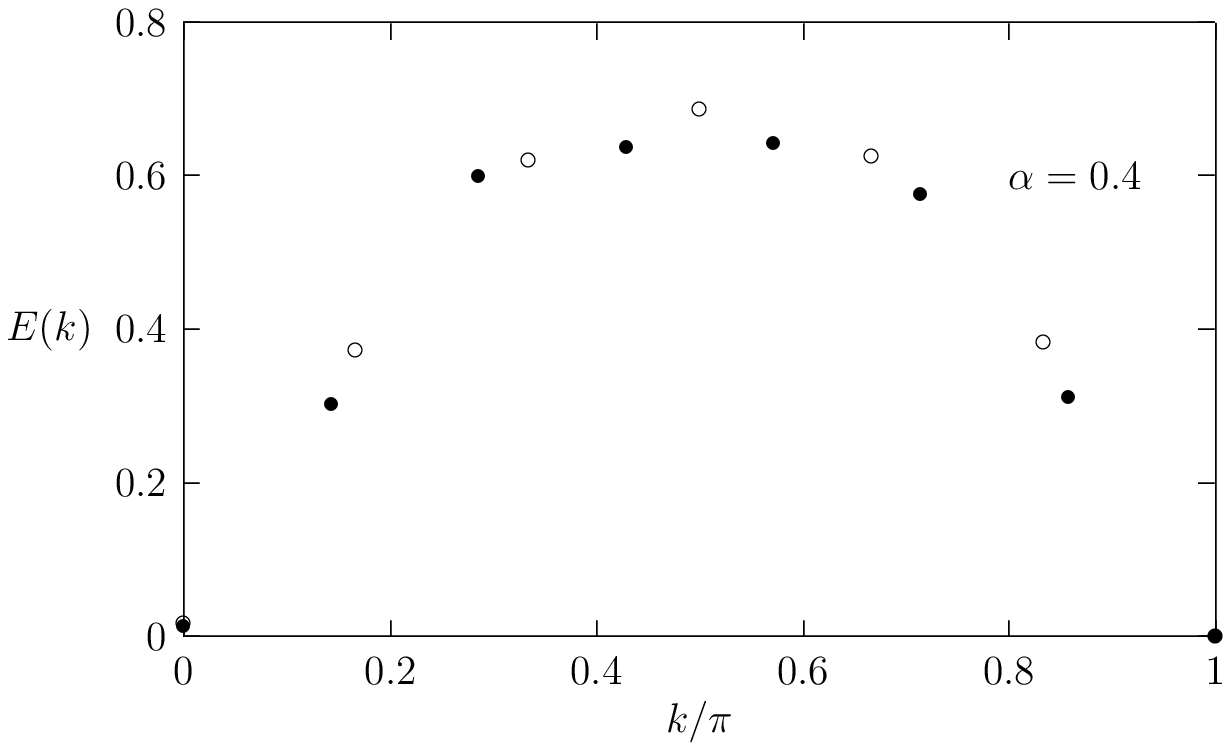}
\caption{\label{fig:ek} ED results for the lowest excited states 
of the system $(s_1,s_2)=(1,\frac{1}{2})$ for two different 
frustration parameters ($\alpha =1$ and $0.4$):  $L=12$ (open circles),
$L=14$ (filled circles). Curves represent  the two
branches of  spin-wave
modes in the paramagnetic Brillouin zone, 
as obtained from the linear spin-wave theory.
The spin-wave energies are multiplied  by the normalization 
factor $v_s/v_{sw}=1.29$ 
where  $v_s=3.87$ and $v_{sw}=2S=3$ are the spin-wave
velocities, as obtained by  the density matrix renormalization group 
method\cite{hallberg} and the linear spin-wave  theory.
Note that the classical transition
point at $\alpha=0.4606$ appears as  an instability point for the
lower spin-wave branch. 
}
\vspace{-.7cm}
\end{center}
\end{figure}

The  above statements concern  the special point $J_2=J_1$ where $S$ is a good 
quantum number. As an example, in Fig.~\ref{fig:ek} we show the energies of 
the lowest excited states ($L=12$ and $14$) for  $\alpha =0.4$
and $1$. Apart from the $k=0$ state, which is characterized by the  total spin
$2$, the lowest excited  states are triplets above the singlet ground state.
This structure of the low-energy spectrum is valid in the whole region
$\alpha >\alpha_{c2}$  up to the limit $\alpha= \infty$  where
the system is composed of two independent $s_1=1$ and $s_2=\frac{1}{2}$
antiferromagnetic Heisenberg chains. In accord to the generalized  
Lieb-Schultz-Mattis (LSM) theorem,~\cite{affleck2} it is natural 
to suppose  that the gapless linear structure of the  spectrum 
around $k=\pi$ survives
away from the  point $J_2=J_1$.

To study the properties of the quantum paramagnetic phase
in the whole region $\alpha>\alpha_{c2}$, we may compare the
finite-size scaling properties of the ground state and the lowest excited
states with those based on the $SU(2)$ Wess-Zumino-Witten  (WZW) nonlinear
$\sigma$ model.
 This  model with the topological coupling $k_0=1$ is believed to describe
the  antiferromagnetic Heisenberg 
chains with half-integer site spins.~\cite{affleck1}. In the following  
we restrict our analysis to the   
$(s_1,s_2)=(1,\frac{1}{2})$ system. 
According to the conformal field theory,
the ground state energy $E_0(L)$ of a periodic system with length $L$
is given by the following expression\cite{affleck1}
\be\label{e0}
\frac{E_0}{L}=\varepsilon_0-\frac{\pi v_s}{6L^2}\left[
1+\frac{3}{8}g^3+{\cal O}(g^4)\right]+\frac{a_1}{L^4} \, .
\ee
Here $\varepsilon_0$ is the ground state energy per rung in the
thermodynamic limit, $v_s$ is the spin-wave velocity,  and 
$g=g(L)$ is the effective coupling constant
of the marginally irrelevant operator $-2\pi g{\bf J}_L\cdot {\bf J}_R$
at the length scale $L$. ${\bf J}_L$ and ${\bf J}_R$ are the conserved current
operators for the left and right movers in the WZW theory. The $L^{-4}$ 
contribution comes from irrelevant operators. The coupling 
$g$ is defined by the  
renormalization-group (RG) equation\cite{solyom,lukyanov} 
\be\label{g}
\frac{1}{g}+\frac{1}{2}\ln g=\ln\frac{L}{L_c}\, ,
\ee
where $L_c$ is a non-universal effective length scale depending on the 
microscopic model. 
An iterative solution of (\ref{g}) yields the following expansion 
for $g$ 
\be\label{exp:g}
g=\frac{1}{\ln(L/L_c)}-\frac{\ln\ln(L/L_c)}{2\ln^2(L/L_c)}+{\cal O}
\left(\frac{1}{\ln^3(L/L_c)}\right)
\ee
so that the marginally irrelevant operator introduces logarithmic
corrections in Eq. (\ref{e0}). 

The energy of the lowest 
triplet excitation $E_t(L)$ with  momentum $k=\pi$ can be expressed in the form
\be\label{et}
\frac{E_t-E_0}{L}=\frac{2\pi v_s}{L^2}\left[
\frac{1}{2}-\frac{g}{4}
+b_1g^2+{\cal O}(g^3)\right]
+\frac{b_2}{L^4} \, .
\ee 
At  moderate length scales ($L$=8,10,12, and 14),  the coupling constants
$g(L)$ in (\ref{e0}) and (\ref{et}) may have  different values, so that
instead of $L_c$ we introduce the effective length scales $L_0$ and $L_t$
for the ground state energy and the energy of
 triplet excitations.   
\begin{figure}[hbt]
\samepage
\begin{center}
\includegraphics[width=8cm]{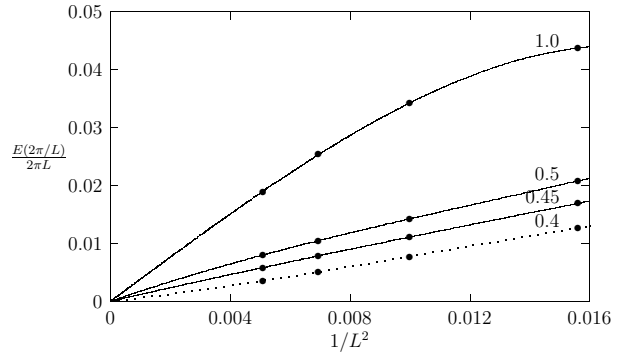}
\caption{\label{fig:vs}  
Reduced energy gap between the 
lowest excited state at $\pi-2\pi/L$     
and the ground state. The interpolation of ED data is performed by
the ansatz $E(2\pi/L)/2\pi L=\sum_{n=1}^3c_nL^{-2n}$. 
For $\alpha=0.45,0.5$, and $1$ this  yields, respectively,
the spin-wave velocities $v_s=1.19,1,78$, and $3.81$. 
Note that the  interpolation function for $\alpha =0.4$
has a positive curvarute.
}
\vspace{-.7cm}
\end{center}
\end{figure}                                      

The parameter $v_s$ in Eqs. (\ref{e0}) and (\ref{et})
can be independently determined from the  scaling of the reduced  
gap $E(2\pi/L)/2\pi L$, where $E(2\pi/L)$ is the energy gap between the 
lowest excited state with momentum  $\pi-2\pi/L$ and the  ground state.
Using the interpolation ansatz  $E(2\pi/L)/2\pi L=\sum_{n=1}^3c_nL^{-2n}$,  
we find, in particular, the estimate 
$v_s\equiv c_1=3.81$ at  $\alpha =1$. This is close to  the density matrix RG  
result $3.87\pm 0.02$
for the antiferromagnetic $S=\frac{3}{2}$ Heisenberg chain.~\cite{hallberg} 
In Fig.~\ref{fig:vs} we  present
the interpolation curves for  different values of the
frustration parameter $\alpha$. Excluding the point $\alpha=0.4$, 
our  estimates for $v_s$  can be well interpolated  
by the ansatz  
\be\label{vs}
v_s=v_1(\alpha-\alpha_{2c})^{\gamma}+v_2(\alpha-\alpha_{2c})
+{\cal O}\left[(\alpha-\alpha_{2c})^2\right]
\ee
up to $\alpha=1$, provided that $\gamma \approx\frac{2}{3}$. 
The linear spin-wave theory gives the exponent
$\gamma=\frac{1}{2}$. Note that the spin-wave ansatz (\ref{vs}) assumes
that the velocity $v_s$ vanishes at the critical point $\alpha_{c2}$.
Of course, the above interpolation of ED data  can not definitely  confirm 
such an  assumption, although  the apparent change in  the curvature 
of $E(2\pi/L)/2\pi L$  $vs$ $1/L^2$ close to  $\alpha =0.4$ gives 
some indication in favor of $v_s(\alpha_{c2})=0$.\cite{cabra}
\begin{figure}[hbt]
\samepage
\begin{center}
\includegraphics[width=8cm]{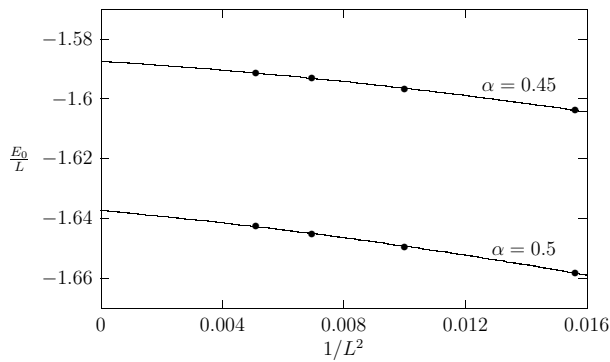}
\caption{\label{fig:e0} 
Scaling of the ground state energy $E_0(L)/L$
at $\alpha=0.45$ and $0.5$. The interpolation  of ED data 
  is performed by Eqs. (\ref{e0}) and (\ref{g}).
}
\vspace{-.7cm}
\end{center}
\end{figure}                                      

Having the parameter $v_s$ for different $\alpha$,
now we can interpolate  the  ED data for  $E_0(L)$ and $E_t(L)$
by using the scaling expressions (\ref{e0}) and (\ref{et}).
The fitting parameters in  (\ref{e0})
are $\varepsilon_0$, $a_1$ and $L_0$. Alternatively, in Eq. (\ref{et}) the
fitting parameters are $b_1$, $b_2$, and $L_t$.
As an example, in Fig.~\ref{fig:e0} we present the interpolation 
curves $E_0(L)/L$ $vs$ $1/L^2$ for frustration parameters
$\alpha=0.45$ and $0.5$. Using the RG equation (\ref{g}) for $g(L)$, the 
best fit  at $\alpha=1$ is obtained for
$\varepsilon_0=-2.3290$, $L_0=0.94$ and $a_1=-20.5$. This is in
accord with the  density matrix RG result 
$\varepsilon_0=-2.32833$.~\cite{hallberg}
The parameter $L_0=0.94$ corresponds to an effective coupling 
constant $g(10)=0.35$. Performing the fits   down to $\alpha =0.4$, we
observe that the characteristic length $L_0$
remains almost unchanged (excluding  
the point $\alpha =0.4$ where formally $L_0\rightarrow \infty$).
An  interpolation procedure using
only the leading term in the logarithmic expansion (\ref{exp:g})
produces  similar results, although with a slightly larger effective 
length $L_0$ ($=1.08$). 
\begin{figure}[hbt]
\samepage
\begin{center}
\includegraphics[width=8cm]{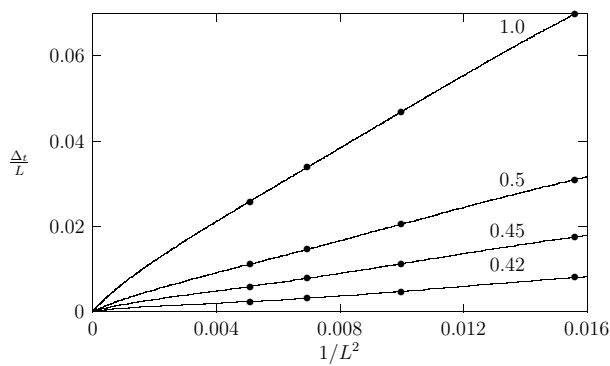}
\caption{\label{fig:dt} 
Scaling of the triplet gap $\Delta_t/L=[E_t(L)-E_0(L)]/L$
for the frustration parameters $\alpha=0.42,0.45,0.5$, and $1$. 
The fit of ED data is performed with  the scaling formula
 (\ref{et}) and  by using the logarithmic expansion 
(\ref{exp:g}) up to second order in $1/\ln (L/L_t)$.
}
\vspace{-.7cm}
\end{center}
\end{figure}                                      
\begin{table}
\caption{\label{tab:table1} Values of the coefficients in  (\ref{e0}) and 
(\ref{et}) giving the best fit to the ED data.
The values of $v_s$ correspond to the ansatz (\ref{vs}) with $\gamma =\frac{2}{3}$
giving the best fit to the  estimates from Fig.~\ref{fig:vs}. 
$L_0=0.94$ for all values of  $\alpha$ presented in the Table. 
}
\begin{ruledtabular}
\begin{tabular}{ccccccc}
$\alpha$ &$v_s$&$\varepsilon_0$
 &$a_1$&$L_t$&$b_1$&$b_2$\\
\hline
0.42& 0.71 & -1.5608 & -26.8
& 2.20 & -2.2 & 318\\
0.45& 1.21 & -1.5873 & -26.0
& 2.20 & -1.8 & 449\\
0.50& 1.77 & -1.6373 & -25.5
& 2.12 & -1.5 & 495\\
1.00& 3.87 & -2.3290 & -20.5
& 2.06 & -1.4 & 942\\
\end{tabular}
\end{ruledtabular}
\end{table}

The  ${\cal O}(g)$ correction to  the scaling dimension in (\ref{et})
makes the  fit  of our ED data  more intricate. 
Moreover, we have found that for  $L\leq 14$  the lowest
singlet excited states  $E_s(L)$ may belong to different
conformal towers. That is why, instead of utilizing the 
combination of $E_t(L)$ and $E_s(L)$ which eliminates the ${\cal O}(g)$
correction,~\cite{ziman} we have  performed the interpolation 
directly with  Eq. (\ref{et}) by using the logarithmic expansion (\ref{exp:g})     
up to second order in $1/\ln(L/L_t)$. The results  presented in
 Fig.~\ref{fig:dt} and Table \ref{tab:table1}  imply that  the effective coupling
at a given length scale $g(L)$ exhibits only a small increase\cite{note2}
when  approaching
the critical point $\alpha_{c2}$, in agreement with the interpolation result
for $E_0(L)$. On the other hand, as $\alpha \rightarrow \alpha_{c2}$
one indicates a monotonous growth of the ${\cal O}(g^2)$ contribution
to the scaling dimension. The  scaling  behavior of $E_0(L)$ and
$E_t(L)$ for $\alpha >1$ qualitatively reveals the same properties. 
\section{Conclusions}
In conclusion, we have examined the impact of magnetic frustration 
on the ground-state phase diagram
of two  coupled mixed-spin $(s_1,s_2)=(1,\frac{1}{2})$  
ferrimagnetic Heisenberg chains. The analysis of ED data implies
an interesting  phase diagram containing the ferrimagnetic phase
and a  singlet  paramagnetic phase exhibiting the
characteristics of the  critical spin-liquid
phase in half-integer-spin  antiferromagnetic Heisenberg chains.
Both  phases are separated by a tiny but finite region occupied
by a critical partially-polarized ferromagnetic phase.   
It is natural to expect similar phase diagrams for the whole class of 
frustrated $(s_1,s_2)$ two-leg ladders with half-integer rung spins.        
\begin{acknowledgments}
A part of  the numerical calculations were  performed with 
the {\em Spinpack} program package created by J. Schulenburg. 
This work was  supported by the Deutsche Forschungsgemeinschaft
(Project No. 436BUL/17/5/03).
\end{acknowledgments}


\begin{thebibliography}{}
\bibitem{note1}  For brevity, we do not present  the ED data revealing
the anisotropy in the spin-spin correlations 
in both magnetic phases.
\bibitem{affleck2} Note that the $n$-th unit cell of the model (\ref{h}) 
 contains only the rung spins ${{\bf s}_1}_n$ and ${{\bf s}_2}_n$
so that for $(s_1,s_2)=(1,\frac{1}{2})$ we can apply the generalized
LSM theorem: see,  Affleck and E. Lieb, 
Lett. Math. Phys. \textbf{12}, 57 (1986).
\end{thebibliography}
\end{document}